\documentclass{iau}
\usepackage{graphicx}
\newcommand{\rpup} {$\rho$~Pup}
\newcommand{\dxcet} {DX~Cet}

\title[Baade-Wasselink projection factor] 
{The Araucaria Project : the Baade-Wesselink projection factor of pulsating stars}

\author[N. Nardetto et al.]   
{Nicolas Nardetto $^1$, Jesper Storm $^2$, Wolfgang Gieren $^3$, Grzegorz Pietrzynski$^4$, Ennio Poretti$^5$
}

\affiliation{$^1$Laboratoire Lagrange, UMR7293, Universit\'e de Nice Sophia-Antipolis, CNRS, Observatoire de la C\^ote d'Azur, Nice, France   \\ email: {\tt Nicolas.Nardetto@oca.eu} \\[\affilskip]
$^2$Leibniz Institute for Astrophysics, An der Sternwarte 16, 14482, Potsdam, Germany \\email: {\tt jstorm@aip.de}\\
$^3$Departamento de Astronom\'ia, Universidad de Concepci\'on, Casilla 160-C, Concepci\'on, Chile \\email: {\tt wgieren@astro-udec.cl}\\
$^4$Warsaw University Observatory, Al. Ujazdowskie 4, 00-478, Warsaw, Poland \\email: {\tt pietrzyn@astrouw.edu.pl}\\
$^5$INAF -- Osservatorio Astronomico di Brera,  Via E. Bianchi 46, 23807 Merate (LC), Italy \\email: {\tt ennio.poretti@brera.inaf.it}

}

\pubyear{2013}
\volume{301}  
\pagerange{}
\jname{Precision Asteroseismology: Celebration of the Scientific Opus of
Wojtek Dziembowski}
\editors{W. Chaplin, J. Guzik, G. Hander \& A. Pigulski, eds.}
\begin{document}

\maketitle

\begin{abstract}
The projection factor used in the Baade-Wesselink methods of determining the distance of Cepheids makes the link between the stellar physics and the cosmological distance scale. A coherent picture of this physical quantity is now provided based on several approaches. We present the lastest news on the expected projection factor for different kinds of pulsating stars in the Hertzsprung-Russell diagram.

\keywords{Stars: oscillations (including pulsations) -- Stars: atmospheres }
\end{abstract}

\firstsection 
\section{Short review on the projection factor of Cepheids}

Since decades the Cepheid stars have been used to calibrate the distance scale and the Hubble constant through their well-know period-luminosity ($PL$) relation (\cite{riess11} and  \cite{freedman10} for a review). Recently,  using the Baade-Wesselink ($BW$) method to determine distances of Cepheids, \cite{storm11a} found that the $K$-band PL relation is nearly universal and can be applied to any host galaxy largely independent of metallicity. The projection factor is a key quantity of the BW methods: it is used to convert the radial velocity variation into the pulsation velocity of the star. There are several ways to study the projection factor. One can use geometrical or static models, hydrodynamical analysis, or even direct observations when the distance of the star is known. In the purely geometric approach, two effects are considered only: the limb-darkening of the star (in the continuum) and the expansion of the atmosphere (at constant velocity). The projection factor is then an integration of the pulsation velocity field (associated with the line-forming region) projected on the line of sight and weighted by the surface brightness of the star, which is defined for instance by $I(\cos(\theta))=1-u_{\mathrm V}+u_{\mathrm V}\cos(\theta)$, where $u_{\mathrm V}$ is the limb darkening of the star in V band and $\theta$ is the angle between the normal of the star and the line of sight \cite{claret11}.  In this case, the geometric projection factor can be derived as follows : $p_\mathrm{0}=\frac{3}{2}-\frac{u_{\mathrm V}}{6} $ \cite{getting34}.  However, this definition of the projection factor implies a specific method of the radial velocity determination, which is the first moment  or centroid method \cite{burki82}. Depending on the limb-darkening considered for Cepheid studied, the value of the projection factor is different: $p=\frac{24}{17}=1.415$ ($u_{\mathrm V}=0.60$, \cite{getting34}), $p=1.375$ ($u_{\mathrm V}=0.75$, \cite{vh52}) or $p=1.360$ ($u_{\mathrm V}=0.80$,  \cite{burki82}).  The latter value has been widely used in spectroscopy. Recently, \cite{neilson12} derived the geometrical projection factor as a function of the period and for several photometric bands using a radiative transfer in spherical geometry and found a  slightly lower value of $p=1.33$ for $\delta$~Cep. Additional studied should be also mentioned like  \cite{gray07} and \cite{hadrava09} in which a geometrical model is directly fitted to the observed spectral line profile (the pulsation velocity is then an output). Such approaches are formally consistent with the geometrical method. 


The second approach to study the Baade-Wesselink projection factor is to consider a hydrodynamical model, which describes the dynamical structure of the atmosphere of the star (in particular atmospheric velocity gradient). Using a so-called {\it piston} model in which the radial velocity curve is used as an input, \cite{s95} found a mean value of the projection factor of $p=1.34$. However, this value was derived using the bi-sector method of the radial velocity determination (applied to theoretical line profiles), which makes the comparison with other studies quite uncertain, unfortunately. On the other hand, using a {\it self-consistent} model of the pulsation (requiring few fundamental parameters such as the stellar mass, the luminosity, the effective temperature and the chemical composition), \cite{nardetto04} found that the atmospheric velocity gradient (and other dynamical effects) reduce the geometric projection factor (found at $p_\mathrm{0}=1.39$ with the model) by about 9\%, leading to a projection factor of $p=1.27\pm0.01$. This value is however consistent with the Gaussian fit method of the radial velocity determination (applied to a spectral line with a typical depth of $D=0.2$). In this 9\% decrease, 5\% comes from the dynamical structure of the atmosphere and 4\% from using the Gaussian fit method. Indeed later,  \cite{nardetto07} provided a revised value of the projection factor, $p=1.33\pm0.02$,  applicable together with the first moment method (and consistent with a plan parallel model atmosphere). It is worth noticing that the projection factor is generally supposed as constant with the pulsation phase following \cite{nardetto06b}.
If one use the cross-correlated radial velocity (which includes many lines and also a Gaussian fit of the cross-correlated mean line profile with a typical depth of $D=0.25$), a lower value of the projection factor is found (of about 11\% compared to the initial geometrical projection factor $p_\mathrm{0}=1.39$), i.e. $p=1.25\pm0.05$ (\cite{nardetto09}). One can say approximatively that in these 11\%, 7\% comes from the dynamical structure of the atmosphere and 4\% from the Gaussian fit.

In 2005, M\'erand et al. applied the {\it inverse} Baade-Wesselink method using the infrared FLUOR/CHARA interferometric observation of $\delta$~Cep. In this approach, the projection factor is fitted, while the distance of $\delta$~Cep is know (from the HST parallax) at the 4\% level (\cite{b82}). They found $p=1.27\pm0.05$ (using the cross-correlation method to derive the radial velocity).  Then, deriving the infrared surface brigthness angular diameters of $\delta$~Cep, and applying again the {\it inverse} BW method, \cite{gro07} and \cite{laney09} found similarly a value of the projection factor of $p=1.27$. Later, \cite{storm11b} constrained {\it directly} the period-projection factor ($Pp$) relation using spectroscopic and photometric observations of Cepheids in the Large Magellanic Cloud (hereafter LMC). In this method, the zero-point of the $Pp$ relation is again based on the HST trigonometric parallaxes of Galactic Cepheids, but the slope is derived from the BW distances of LMC Cepheids (all Cepheids in the LMC used by Storm can be assumed to be at the same distance, leading to an extra constraint on the period projection factor relation). The corresponding value for $\delta$~Cep itself is $p=1.41\pm 0.05$. It has been shown that the metallicity has no impact (at least theoretically) on the projection factor \cite{nardetto11}. Using a similar method \cite{gro13}  found recently a value of the projection factor which is also quite high ($p=1.33$). The latest result comes from \cite{pilecki13}, who constrained the projection factor using a short-period Cepheid ($P=3.8$ days) in a eclipsing binary system. They found $p=1.21\pm0.04$. 

This short review shows that a lot of work has been done to constrain the BW projection factor. And even if some discrepancies remain concerning the inverse photometric BW method of determining the projection factor, a consensus is currently emerging. In particular, we emphasize that the fact that the projection factor derived from the surface-brightness technics is overestimated has no impact on the distances, because at the same time, the amplitude of the photometric angular diameter curve is underestimated. One can say finally that the photometric version of the BW method is {\it self-consistently} calibrated using the HST parallaxes to set the zero point and the distances to LMC Cepheids 
with a large range of periods to  constrain the p-factor relation with pulsation period. However,  \cite{ngeow12} found indeed a significant dispersion in the period-projection factor relation, and this should be also investigated. 

\section{The projection factor for other types of pulsating stars}

\begin{figure}
   \centering
   \includegraphics[width=7cm,height=7cm]{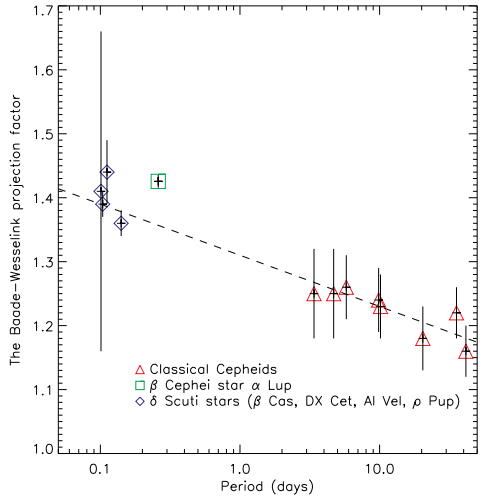}
   \caption{The Baade-Wesselink projection factor as a function of the period for different kinds of pulsating stars. The $\delta$~Scuti stars indicated as blue diamonds are, by increasing period: $\beta$~Cas, DX~Cet, AI~Vel and $\rho$~Pup. The case of the $\beta$ Cephei stars $\alpha$~Lup is described in \cite{nardetto13b}  }
              \label{Pp_pulsating}
    \end{figure}

One possible way to better understand the dynamical structure of Cepheids, and in particular the k-term (\cite{nardetto06a, nardetto08a}), the mass loss (\cite{nardetto08b}), and the projection factor is to perform comparison with other kinds of pulsating stars (as soon as they pulsate in a dominant radial model).

In the framework of the Araucaria Project (\cite{gieren05}) of distances determination in the Local Group, we determined the Baade-Wesselink 
projection factor for four $\delta$~Sct stars: \rpup\, ($p=1.36\pm0.02$), \dxcet\, ($p=1.39\pm0.02$), AI~Vel  ($p=1.44\pm0.05$), and $\beta$~Cas ($p=1.41\pm0.25$). 
Refer to \cite{nardetto13a} for  \rpup\ and  \dxcet\  and to \cite{guiglion13} for AI Vel and $\beta$~Cas. 
Figure~\ref{Pp_pulsating} shows how all these values fit in  an excellent way the extension toward short periods of the relation found for Cepheids,
i.e., $p = [-0.08 \pm 0.05] \log P + [1.31 \pm 0.06]$ \cite{nardetto09}. This result seems more robust than the similar one obtained by \cite{laney09} using an indirect method
based on the comparison of geometric and pulsation parallaxes.
On the other hand, the  projection factor  of the $\beta$~Cep star $\alpha$~Lup is 8$\sigma$ above the relation 
(Fig.~\ref{Pp_pulsating}). By omitting $\alpha$ Lup we can find a common relation to
$\delta$ Sct stars and classical Cepheids. 

\section{Conclusion}

The projection factor is a very complex quantity which involves all the physical structure of the Cepheids' atmosphere. Nevertheless, it is now well constrained using geometrical, hydrodynamical modelling and also direct observations (trigonometric parallaxes and interferometry). Thanks to these efforts to better understand the projection factor, the BW technique of distance determination is becoming one the more robust method in the path to the Hubble constant.

\end{document}